\documentclass[11pt]{fizik}

\usepackage{hyperref}
\hypersetup{
colorlinks=true,
urlcolor=blue,
citecolor=blue}
\usepackage[all]{xy,xypic}
\usepackage{amsfonts,amssymb,amsmath,amsgen,amsopn,amsbsy,theorem,graphicx,epsfig}
\usepackage{eufrak,amscd,bezier,latexsym,mathrsfs,enumerate,multirow}
\usepackage[utf8]{inputenc}
\usepackage[english]{babel}
\usepackage[dvipsnames]{xcolor}
\usepackage{academicons}
\definecolor{orcidlogocol}{HTML}{A6CE39}
\usepackage[pagewise]{lineno}
\usepackage{epstopdf}
\usepackage{xspace}
\usepackage{color}
\usepackage{units}
\usepackage{slashed}
\usepackage{gensymb}
\usepackage{booktabs}
\usepackage{xcolor}
\usepackage{setspace}
\usepackage{physics}
\usepackage{nicefrac}
\usepackage{bbold}
\usepackage{accents}
\usepackage{mathtools}

\def\bt{\begin{equation}}
\def\bea{\begin{eqnarray}}
\def\ee{\end{equation}}
\def\eea{\end{eqnarray}}

\yil{}
\vol{}
\fpage{}
\lpage{}
\doi{}

\title{
Rotating anisotropic Bose gas with large number of vortices}            
\author[KELE\c{S}]{                                                                                     
\textbf{Ahmet Kele\c{s}\thanks{akeles@metu.edu.tr}}\\ 
Department of Physics, Middle East Technical University, Ankara, Turkiye \\               
\\ [1.8em]

\rec{.. 2022}
\acc{.. 2022}
\finv{.. 2022}
}

\newcommand{\bc}{\begin{center}}
\newcommand{\ec}{\end{center}}

\renewcommand{\phi}{\varphi}

\begin{document}

\maketitle

\begin{abstract} 
Rapidly rotating atomic gases provide a platform for studying phenomena akin to type-II superconductors and quantum Hall systems. 
Recently, these systems have attracted renewed interest due to technological advances in the trap anisotropy control, in-situ observation capabilities, and cooling and rotating complex atomic species such as dipolar gases. Understanding the vortex lattice formation and quantum melting is crucial for exploring quantum Hall physics in these systems. In this paper, we theoretically investigate the vortex lattices in anisotropic quantum gases. We formulate the rotating gas Hamiltonian in the Landau gauge, and consider the effects of additional perturbations such as the trap potential in the lowest Landau level (LLL). Focusing on the gases with short-range interactions, we obtain the many-body Hamiltonian projected to lowest Landau level. We consider the limit of full Bose-Einstein condensation and obtain the governing Gross-Pitaevskii equation to identify the possible vortex phases. We numerically solve the Gross-Pitaevskii equation using the imaginary time evolution, and demonstrate the possible vortex lattices as a function of anisotropy, rotation speed and interaction strength. We show that the number of states with a support in the LLL, which determines the number of vortices, follows a Thomas-Fermi type scaling albeit with slightly different coefficients from the usual condensates.
\end{abstract}

\section{Introduction}
\label{section1} 

After the realization of atomic gas Bose-Einstein condensates (BECs) in 1995, ultracold atomic gases emerged as a powerful platform for simulating condensed matter phenomena. One of the earliest and most influential frontiers in this field was the study of the rotating atomic gases, which mimic phenomena well established in type-II superconductors and quantum Hall systems \cite{fetter2009}. Since atomic gases are electrically neutral, quantum mechanical gauge fields cannot be induced using external magnetic fields. Instead, mechanical rotation serves as an alternative, based on the fact that the Coriolis force in the rotating frame affects particles in the same way the Lorentz force acts on charged particles.
%
BECs consist of a macroscopic occupation of a complex wavefunction characterized by a norm and a phase, therefore the influence of rotation or stirring, which breaks time-reversal symmetry explicitly, is completely different from classical fluids. The requirement of the single-valuedness of the condensate wavefunction enforces irrotationality in the system, i.e. line integral of superfluid velocity---which is the gradient of superfluid phase---along any closed contour is zero. However, the gas can still accommodate circulation or angular momentum through the formation of singular vortex punctures \cite{landau2013statistical}. When the rotation speed increases, or equivalently, when the angular momentum of the cloud grows, the number of vortices rises dramatically, eventually forming an array of singularities within the gas \cite{stringari2003}. In the limit of rapid rotation, which will be made more precise below, the system enters into the so-called mean-field quantum Hall regime, where a dense triangular lattice of vortices emerges \cite{ho2001,mueller2003,aftalion2005}. For even faster rotation and higher vortex density, the cores of the neighboring vortices start to coalesce, number of vortices approach to the number of particles, and the resulting strong quantum fluctuations are predicted to melt the vortex lattice, paving the way for more intriguing correlated phases such as fractional quantum Hall states \cite{cooper2001,macdonald2002}.
A series of remarkable experiments have observed atomic gases containing hundreds of vortices, with early indications of vortex lattice melting \cite{cornell2004, dalibard2004}. However, a major challenge in this approach was the vanishing of trapping potential in the fast-rotation limit due to centrifugal forces, leading to significant atom loss \cite{viefers2008,cooper2008}. The rotation, which enables an analogy with the magnetic field due to the Coriolis force, also dismantles the vapor at ultrafast speeds due to centrifugal force. This limitation ultimately motivated the development of alternative schemes for realizing synthetic gauge fields and synthetic dimensions that can probe LLL physics \cite{chalopin2020} in ultracold atomic systems (For review see \cite{cooper2012,Zhai2015,spielman2019}).

Recently, rapidly rotating ultracold systems have gained renewed interest due to technological advances in precise control of the trap anisotropy in rotating gases, in-situ observation capabilities, as well as new techniques in trapping and cooling more complicated atomic species. In Ref.~\cite{zwierlein2021}, using an anisotropic trapping, the rotation frequency was smoothly ramped up to a value above the smaller trap frequency but below the larger one, creating a saddle potential. This enabled the distillation of a single Landau level wavefunction so that the atomic vapor was squeezed down to the scale of a cyclotron orbit in one direction while extending in the other to such an extent that the interparticle distance became comparable to the lowest Landau level (LLL) cyclotron orbit. This opened a new avenue for pursuing strongly correlated phenomena. Subsequent experiments indeed demonstrated broken translational invariance due to interparticle interactions, as well as in-situ observation of the dynamics of the edge modes \cite{zwierlein2022,zwierlein2023}.

In a completely different front, vortex states were realized in dipolar atomic gases experimentally, based on the so-called magnetostriction and magnetostirring of the system \cite{ferlaino2021,ferlaino2022}.
These are atomic gases with strong magnetic dipole moments, where all dipoles in the gas can be oriented in the same direction by an external magnetic field. Whereas the confinement is still fully isotropic, due to anisotropic dipole-dipole interaction which favors head-to-tail alignment of the dipoles, the cloud elongates along the tilting direction, which is typically a few percent depending on the tilting angle. However, rapid rotation can dramatically amplify this small anisotropy in a similar manner to gases in anisotropic traps \cite{fetter2001, oktel2004,fetter2007}. Recently, such dipolar BECs have demonstrated the possibility of phase coherence coexisting with broken translational invariance which was manifest through vortices \cite{ferlaino2023,ferlaino2024}.

Motivated by these remarkable advances, in this paper, we focus on the theory of BECs in the LLL and investigate resulting vortices and vortex lattices. After formulating the problem of the rapidly rotating atomic gas in the Landau gauge, and demonstrating the analogy with charged particles under magnetic field, we derive single particle equations with additional perturbations such as trap potential in the LLL. Starting from a second quantized form, we derive the full many-body Hamiltonian in the LLL. To demonstrate vortices as a function of interactions strength, trap anisotropy, rotations rate and number of particles, we go to the limit of complete condensation and derive a Gross-Pitaevskii equation in the LLL. We develop a robust solution of the Gross-Pitaevskii equation based on imaginary-time evolution and show transitions between different vortex lattice structure as a function of anisotropy and interactions strength.

This paper is organized as follows: In Sec.~\ref{section2}, we formulate the rotating gas problem in the single-particle limit and demonstrate the transition to the Landau gauge. We also examine the effects of single-particle perturbations, such as the trap potential, and calculate observables, including angular momentum. In Sec.~\ref{section3}, we develop a second-quantization formalism for the many-body problem with short-range interactions, reducing it to the lowest Landau level. In Sec.~\ref{section4}, we explore the full Bose-Einstein condensation limit, derive the Gross-Pitaevskii formalism in the lowest Landau level, and solve it via imaginary time evolution to reveal the ground-state vortex structures. Finally, in Sec.~\ref{section5}, we conclude with a summary of our findings and discuss future research directions.

\section{Single Particle Hamiltonian in the Landau Gauge}
\label{section2} 

Single particle Hamiltonian of a trapped rotating Bose gas in the rotating frame can be written as \cite{fetter1999}
\begin{equation}
\label{equation1}
    H = \frac{p_x^2+p_y^2}{2M}+\frac{1}{2}M\omega_x^2x^2
    +\frac{1}{2}M\omega_y^2y^2-\Omega(xp_y-yp_x),
\end{equation}
where $M$ is atomic mass,  $p_{x}=-i\hbar\partial_{x}$ (same for $y$) is the momentum operator with commutation $[x, p_x]=i\hbar$, $\omega_{x}$ and $\omega_y$ are frequencies of the trap potential in $x$ and $y$ directions, and $\Omega$ is the mechanical rotation frequency of the cloud. The last term in the parenthesis is the angular momentum operator along the direction perpendicular to two dimensional gas.
For asymmetric trap potentials with broken rotational symmetry, it is convenient to take $x$ and $y$ directions in unequal footing, which suggests the use of the Landau gauge. Let us, however, start with the symmetric gauge which has the vector potential
$\mathbf{A}=-M\Omega\mathbf{r}\times\hat z=m\Omega(-y,x,0)$, which is more widely used in the literature of rotating gases. The following expression can be shown simply by using the definition of vector potential in the symmetric gauge
\begin{equation}
    \frac{(\mathbf{p}-\mathbf{A})^2}{2M} = \frac{\mathbf{p}^2}{2M}+\frac{1}{2}M\Omega^2(x^2+y^2)-\Omega L_z.
\end{equation}
Using this result in Eq.~\eqref{equation1}, the Hamiltonian in an asymmetric trap can be written as
\begin{equation}
    H = \frac{(\mathbf{p}-\mathbf{A})^2}{2M} + 
    \frac{1}{2}M(\omega_x^2-\Omega^2)x^2+\frac{1}{2}M(\omega_y^2-\Omega^2)y^2.
\end{equation}
This result famously demonstrates equivalence of rotation with magnetics field which also yields weakening of the trapping due to centrifugal force. Additional transformation can be obtained from the substitution $\psi(\mathbf{r})=e^{\frac{i}{\hbar}\chi(\mathbf{r})}\psi'(\mathbf{r})$ in the time-independent Schrodinger equation, $H\psi=\varepsilon\psi$, as
\begin{equation}
    \frac{(\mathbf{p}-\mathbf{A})^2}{2M}\psi + V(\mathbf{r})\psi=\varepsilon\psi 
    \rightarrow
    \frac{(\mathbf{p}-\mathbf{A}')^2}{2M}\psi' + V(\mathbf{r})\psi'=\varepsilon\psi', 
\end{equation}
where $\mathbf{A}'=\mathbf{A}-\nabla\chi$. Choosing 
$\chi(\mathbf{r})=-M\Omega xy$ gives $\mathbf{A}'=M\Omega(0,2x,0)$ which is the Landau gauge as we desired earlier. Dropping the primes in the following, the Hamiltonian becomes \cite{Shlyapnikov2005}
\begin{equation}
    H = 
    \frac{p_x^2}{2M}+\frac{(p_y-2M\Omega x)^2}{2M}
    +\frac{1}{2}M(\omega_x^2-\Omega^2)x^2
    +\frac{1}{2}M(\omega_y^2-\Omega^2)y^2.
    \label{eq:H_single_particle}
\end{equation}
We are assuming an anisotropic system with 
 $\Omega\le\omega_y<\omega_x$ and fast rotation $\Omega\rightarrow\omega_y$ such that the gas is elongated along the $y$ direction.
The parameter regime $\omega_y<\Omega<\omega_x$ corresponds to the so-called geometric squeezing in Ref.~\cite{zwierlein2021}.
When $\Omega=\omega_y$, $H$ becomes translationally invariant in $y$ direction. This suggests splitting the Hamiltonian as $H=H_0+V_{trap}(y)$ where 
\begin{align}
    H_0&=\frac{p_x^2}{2M}+\frac{(p_y-2M\Omega x)^2}{2M}
    +\frac{1}{2}M(\omega_x^2-\Omega^2)x^2\\
    V_{trap}(y)&=\frac{1}{2}M(\omega_y^2-\Omega^2)y^2
    \label{eq:Vho}
\end{align}

\begin{figure}[h!]
\centering
\includegraphics[width=7cm]{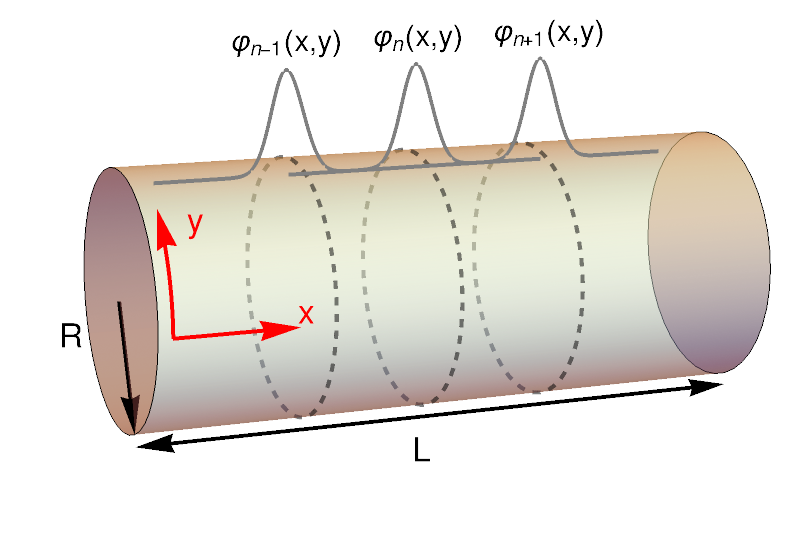}
\caption{Cylindrical geometry of the Landau gauge and a few representative lowest Landau Level wavefunctions. A two dimensional system is considered with dimensions $L$ and $2\pi R$ where the system has periodic boundary conditions along the circumference and open boundaries along longitudinal direction.}
\label{figure1}
\end{figure}
\subsection{Translationally invariant limit along $y$-axis $\omega_y=\Omega$}
We first focus on $H_0$ and substitute solutions of the form $e^{iky}$.
Then $H_0$ can be put in a form identical with a harmonic oscillator by ``completing the square''
\begin{equation}
    \label{equation8}
    H_0=\frac{p_x^2}{2M}+\frac{1}{2}M\omega_c(x-x_k)^2 + \frac{\hbar^2k^2}{2M_*},
\end{equation}
where
\begin{equation}
\omega_c=\sqrt{3\Omega^2+\omega_x^2},\quad 
\frac{M}{M_*}=1-\frac{4\Omega^2}{\omega_c^2}, \quad
x_k=\frac{2\Omega}{\omega_c}k\ell^2, \text{ and } \ell^2=\frac{\hbar}{M\omega_c}.
\end{equation}
The first two terms in \eqref{equation8} are the familiar Harmonic oscillator Hamiltonian, thus the lowest eigenvalue of $H_0$ is given as
\begin{equation}
    \varepsilon_{k} = \frac{\hbar\omega_c}{2} + \frac{\hbar^2k^2}{2M_*}.
    \label{eq:energy_LLL}
\end{equation}
which is the lowest Landau level.
Assuming periodic boundary conditions along $y$ direction, we take a cylinder with radius $R$ and length $L$ as shown in Fig.~\ref{figure1}. Using $y\in[0,2\pi R)$, the wave-numbers are enumerated as $k=\frac{1}{R}n$ with $n\in\mathbb{Z}$ and the corresponding normalized eigenfunction is given by
\begin{equation}
    \phi_{k}(x,y) = 
    \frac{1}{\sqrt{2\pi R}}e^{iky}
    \frac{1}{(\pi\ell^2)^{1/4}}e^{-\frac{1}{2\ell^2}(x-x_k)^2}.
    \label{eq:lll_wavefunction}
\end{equation}
 To find the allowed values of integers $n$, we require the guiding centers $x_k$ to be inside ``the sample domain'' $-L/2\le x_k\le L/2$, which gives
\begin{equation}
    -\frac{N_v}{2}\le n \le \frac{N_v}{2},
    \label{eq:n_summation}
\end{equation}
or $n\in\mathbb{Z}_{N_v}$. 
%
Note that the ``sample domain'' $L$
is treated as a parameter for simplicity. However, as we show below, it is determined by the equilibrium properties of the gas and follows a Thomas-Fermi scaling, similar to Ref.~\cite{ho2001}.
The number of vortices denoted by $N_v$ is defined by
\begin{equation}
N_v=\frac{A}{2\pi\ell^2}\frac{\omega_c}{2\Omega}.
\end{equation}
where $A=L2\pi R$ is the surface area of the cylinder. 
In the isotropic and ultrafast rotation limit $\Omega=\omega_x=\omega_y$ this takes the more familiar form
$N_v\rightarrow A/2\pi\ell^2$.
Roughly, equating the flux trapped in cyclotron orbit to be equal to flux quantum $\Phi_0\equiv h/e_*$ gives a definition for effective magnetic field $B_*$ and effective charge $e_*$ as 
$B_* 2\pi\ell^2 = h/e_*$ such that 
\begin{equation}
    e_*B_*=M\omega_c,
\end{equation}
which agrees with Ref.~\cite{cooper2005} Eq. 17 in the fast rotation limit with isotropic trap $\Omega=\omega_x=\omega_y$. Furthermore, if we define the vortex density as $n_v=N_v/A$, we obtain 
\begin{equation}
n_v=\frac{M\omega_c}{2\pi\hbar}\frac{\omega_c}{2\Omega},
\end{equation}
which takes the same form as the Feynman expression \cite{feynman1955} $n_v=M\Omega/\pi\hbar$ in the limit $\Omega\rightarrow\omega_x=\omega_y$. 
We define the ``anisotropy ratio'' \cite{haldane1994}
\begin{equation}
    \gamma = \frac{\ell}{2\pi R}\frac{2\Omega}{\omega_c},
    \label{eq:gamma}
\end{equation}
the guiding centers can be written as $x_k/\ell=2\pi\gamma n$. When $\gamma$ is small, the widths of the Gaussians in the LLL coalesce, effectively resulting in a one dimensional system along the length of the cylinder. When $\gamma$ is large, neighboring LLL wavefunction completely decouple, forming one-dimensional isolated stripes along the circumference of the cylinder. 
%
The system can be described in terms of the independent variables $\gamma$, $N_v$, and the system size, which is given by  
$L/\ell=2\pi\gamma N_v$. 
In the isotropic limit, where $2\Omega/\omega_c=1$ and assuming $2\pi R=L$, the relation between vortex number $N_v$ and the ratio $\gamma$ is $N_v=2\pi/\gamma^2$. It should be born in mind that even in this limit, the system does not fully correspond to the symmetric gauge limit due to inherent anisotropy in the boundary conditions. Consequently, the ground state and the topological degeneracies remain distinct from those in other geometries \cite{haldane1994}.

\subsection{Harmonic confinement along $y$}
Inclusion of a harmonic confinement $V(y)$ breaks translational invariance along $y$ direction and creates off-diagonal terms in momentum $k$. Let us start with the time independent Schrodinger equation:
\begin{equation}
    \left[H_0 + V_{trap}(y)\right]\psi(x,y) = E\psi(x,y),
    \label{eq:H0_plus_Vx}
\end{equation}
where $V(y)$ is given in Eq.~\eqref{eq:Vho}. We are interested in the physics of LLL, therefore use the expansion
\begin{equation}
    \psi(x,y) = \sum_{k'} c_{k'}\varphi_{k'}(x,y),
    \label{eq:LLL_expansion}
\end{equation}
with $\psi_k(x,y)$ given in Eq.~\eqref{eq:lll_wavefunction}. We multiply and integrate Eq.~\eqref{eq:H0_plus_Vx} with $\int_{0}^{2\pi R}\frac{dy}{2\pi R}e^{-iky}$, which gives
\begin{equation}
    \varepsilon_k c_k + \sum_{k'} V_{trap}(k,k')c_{k'} = E c_k,
\end{equation}
where for $k=\frac{1}{R}n$ and $k'=\frac{1}{R}m$ we get
\begin{equation}
    V_{trap}(n,m) = \frac{1}{2}M(\omega_y^2-\Omega^2)
    R^2e^{-\pi^2\gamma^2(n-m)^2} \times
    \begin{cases}
    2\pi\frac{(-1)^{n-m}}{(n-m)^2}
    &\text{ if } n\neq m,\\
     \frac{\pi^2}{3} &\text{ if } n=m.
    \end{cases}
    \label{eq:harmonic_confinement}
\end{equation}
To obtain this result, we first evaluate the $x$-integral by expanding the domain to $x\in(-\infty,\infty)$, justified by the rapidly decaying Gaussian form. We assume that both $x_k$ and $x_{k'}$ lie within the sample domain and then compute the elementary $y$-integral exactly.

\subsection{Angular momentum in the lowest Landau level}

Due to broken rotational invariance, angular momentum is not conserved and the operator $L_z=xp_y-yp_x$ mixes different wavefunctions in the LLL manifold. However, we can still consider the expectation value of $L_z$ in the Landau gauge wavefunctions, which is useful to make comparison with the symmetric gauge. Expectation value of angular momentum is defined as
\begin{equation}
    \langle L_z\rangle =
    \int_{-L/2}^{L/2} dx
    \int_{-\pi R}^{\pi R} dy
    \psi^*(x,y)
    [
    -xi\hbar\frac{\partial}{\partial y}
    +yi\hbar\frac{\partial}{\partial x}
    ]
    \psi(x,y).
\end{equation}
Using $\psi(x,y)$ is obtained from Eq.~\eqref{eq:LLL_expansion} and Eq.~\eqref{eq:lll_wavefunction} we write
\begin{align}
    \langle L_z\rangle =
    \frac{i\hbar}{\ell^22\pi R}\frac{1}{\sqrt{\pi\ell^2}}
    \sum_{k',k}c_{k'}^*c_k \int_{-L/2}^{L/2}dx
    \int_{-\pi R}^{\pi R}&dy
    e^{i(k-k')y-\frac{1}{2\ell^2}[(x-x_k)^2+(x-x_{k'})^2]}\nonumber\\
    &\times\left[
        x\hbar k -yi\hbar \frac{(x-x_k)}{\ell^2}
    \right],
\end{align}
and the calculation of the integrals gives
\begin{equation}
    \langle L_z\rangle = 
    \hbar\frac{2\Omega}{\omega_c}\sum_n
    \left(\gamma^2n^2+\frac{1}{2}\right) 
    c_n^*c_n,
\end{equation}
where we have taken $L\rightarrow\infty$ limit for analytical calculation as before, and the summation is over values given in Eq.~\eqref{eq:n_summation}.



\section{Second quantized Hamiltonian}
\label{section3}
For a dilute gas that interact with short-range contact interactions, many-body Hamiltonian is written in second quantized form
\begin{equation}
    H = \int d^2r \hat\Psi^\dagger(\mathbf{r})H(\mathbf{r})\hat\Psi(\mathbf{r})+
    \frac{1}{2}g
    \hat\Psi^\dagger(\mathbf{r})
    \hat\Psi^\dagger(\mathbf{r})
    \hat\Psi(\mathbf{r})
    \hat\Psi(\mathbf{r})
\end{equation}
where for a quasi-two-dimensional system tightly confined along perpendicular direction, the coupling constant can be written in terms of the oscillator length normal to 2D plane $a_z$ and s-wave scattering length $a_s$  as 
$g=\sqrt{8\pi}\hbar^2a_s/Ma_z$ \cite{fetter2009}.
$H(\mathbf{r})$ is the single particle Hamiltonian given in Eq.~\eqref{eq:H_single_particle}, $\Psi(\mathbf{r})$ and $\Psi^\dagger(\mathbf{r})$ are bosonic field operators with commutations $[\hat\Psi(\mathbf{r}),\hat\Psi(\mathbf{r}')]=\delta(\mathbf{r}-\mathbf{r}')$. As before we are interested in the LLL, therefore expand the field operators as
\begin{equation}
    \hat\Psi(\mathbf{r}) = \sum_k \hat c_k \varphi_k(x,y),
\end{equation}
where different from the case above here $\hat c_k$ and $\hat c_k^\dagger$'s are bosonic field operators with commutations $[\hat c_k,\hat c_{k'}^\dagger]=\delta_{kk'}$ and they annihilate or create a particles in the LLL labeled by $k\equiv \frac{n}{R}$, for $n\in\mathbb{Z}$ consistent with the boundary conditions. Therefore, in the LLL, we obtain the following second quantized Hamiltonian
\begin{equation}
    H = \sum_n \varepsilon_{n}
    \hat c^\dagger_{n} \hat c_{n}
    +\sum_{n,m}V_{trap}(n,m)
    \hat c^\dagger_{n} \hat c_{m} + 
    \frac{1}{2}\sum_{n,m,l} V_{int}(n,m,l)
    \hat c^\dagger_{n}
    \hat c^\dagger_{m}
    \hat c_{l}
    \hat c_{n+m-l},
    \label{eq:H_many-body_LLL}
\end{equation}
where $\varepsilon_n$ is given in Eq.~\eqref{eq:energy_LLL} with $k=\frac{n}{R}$, $V_{trap}(n,m)$ is given in Eq.~\eqref{eq:harmonic_confinement} and interaction matrix elements are calculated as
\begin{align}
    V_{int}(n,m,l) &= 
    \frac{1}{2\pi R}\frac{g}{\sqrt{2\pi}\ell}
    \exp\left(
    -2\pi^2\gamma^2
    \left[
        (n-l)^2+(m-l)^2
    \right]    
    \right).
\end{align}
This result is also obtained by taking integrals in the limit $L\rightarrow\infty$. The parameter $\gamma$ is defined in Eq.~\eqref{eq:gamma} as the anisotropy parameter of the cloud. This result is consistent with \cite{jolicoeur2012} and also agrees with \cite{Shlyapnikov2005} in the limit $\Omega\rightarrow \omega_y$. An alternative re-labelling,
as presented in \cite{seidel2013}, can also be useful for numerical implementations. We refer to the limit $1/M_*\rightarrow 0 $ and $V_{trap}(n,m)
\rightarrow 0$ as the flat-band limit, where only two-body interactions remain, as seen in the last term of Eq.~\eqref{eq:H_many-body_LLL}. This scenario is particularly relevant for studying correlated states in the literature. Using the coupling constant $g_{2D}$ in quasi-two dimension, the interaction in units of Landau level spacing $\hbar\omega_c$ can be expressed as
\begin{equation}
    \frac{1}{2\pi R}\frac{g}{\sqrt{2\pi}\ell}\frac{1}{\hbar\omega_c}
    = \frac{1}{\pi}\frac{\ell}{ R} \frac{a_s}{a_z}.
\end{equation}
However, in the limit of large Landau level spacing, $\hbar\omega_c\rightarrow\infty$, which is more relevant from an experimental perspective, the key interplay occurs between the kinetic energy (given in the first term of Eq.~\ref{eq:H_many-body_LLL}) and the interaction term.
This suggests using the single particle energy $\hbar^2/2M_*R^2$ in Eq.~\eqref{eq:energy_LLL} as an alternative energy scale 
\begin{equation}
    \frac{1}{2\pi R}\frac{g}{\sqrt{2\pi}\ell}\frac{1}{\frac{\hbar^2}{2M_*R^2}}=\frac{1}{\pi}\frac{R}{\ell}
    \frac{M_*}{M}\frac{a_s}{a_z}\equiv g_*.
\end{equation}
Here we drop the zero point energy of LLL, $\hbar\omega_c/(\hbar^2/2M_*R^2)$, and the trap potential to obtain
\begin{equation}
    H_* = \sum_n n^2 \hat c_n^\dagger \hat c_n +
    \frac{g_*}{2}
    \sum_{n,l,m}e^{-2\pi^2\gamma^2[(n-l)^2+(m-l)^2]}
    \hat c^\dagger_{n}
    \hat c^\dagger_{m}
    \hat c_{l}
    \hat c_{n+m-l}.
    \label{eq:two_parameter_Hamiltonian}
\end{equation}
Hamiltonian $H_*$ has the advantage of being dependent only on two parameters $g_*$ and $\gamma$.

\section{LLL condensate in the Gross-Pitaevskii limit}
\label{section4}
When the occupation of each mode in the LLL is large, one can ignore the commutation relations of the field operators $\hat c_n$ and $\hat c_n^\dagger$ and treat them as complex numbers $\hat c_{n},\hat c_{n}^\dagger\rightarrow c_{n},c_{n}^*$. Then the Hamiltonian operator becomes an energy functional $H\rightarrow E[c_{n},c_{n}^*]$ given as follows
\begin{equation}
    E[c_{n},c_{n}^*] = \sum_n \varepsilon_{n}
    c^*_{n} c_{n}
    +\sum_{n,m}V_{trap}(n,m)
    c^*_{n} c_{m} + 
    \frac{1}{2}\sum_{n,m,l} V_{int}(n,m,l)
    c^*_{n}
    c^*_{m}
    c_{l}.
    c_{n+m-l}.
    \label{eq:GP_energy_functional}
\end{equation}
The problem reduces to the minimization of the energy functional by varying a set of complex numbers $c_n$, $c_n^*$ subject to the constraint
\begin{equation}
    N = \sum_n |c_n|^2,
\end{equation}
where $N$ is the number of atoms in the cloud. Taking the variations of $\delta(E-\mu N)/\delta c_{n}^*$ with respect to $c_n^*$, where the Lagrange multiplier $\mu$ corresponds to the chemical potential, we get the following Gross-Pitaevskii equation \cite{pethick_smith}
\begin{equation}
    \varepsilon_{n}c_{n} + 
    \sum_{m} V_{trap}(n,m) c_{m} +
    \sum_{m,l}V_{int}(n,m,l)
    c_{m}^*c_{l}c_{n+m-l} =
    \mu c_{n},
    \label{eq:time-indep_GPE}
\end{equation}
which is a non-linear time-independent Schrodinger equation projected to the LLL. By adding lowest possible time derivatives in Eq.~\eqref{eq:GP_energy_functional}, one can define effective action in the lowest Landau level and the corresponding equation of motion can be obtained which will give the time dependent Gross-Pitaevskii equation in the lowest Landau level as follows
\begin{equation}
    i\frac{\partial c_{n}}{\partial t} = 
    \varepsilon_{n}c_{n} + 
    \sum_{m} V_{trap}(n,m) c_{m} +
    \sum_{m,l}V_{int}(n,m,l)
    c_{m}^*c_{l}c_{n+m-l}.
    \label{eq:time-dep_GPE}
\end{equation}
This reproduces Eq.~\eqref{eq:time-indep_GPE} by substitution $c_{n}(t)=e^{-i\mu t}c_{n}$. Let us drop the first two terms by considering LLL limit, rescale $c_{n}\rightarrow c_{n}/\sqrt{N}$ and transform to the imaginary time $t\rightarrow -i\tau$. Then the Eq.~\eqref{eq:time-dep_GPE} becomes
\begin{equation}
    \frac{\partial c_{n}}{\partial \tau} = 
    n^2 c_n+
    g_*N
    \sum_{m,l}
    e^{
    -2\pi^2\gamma^2
    \left[
        (n-l)^2+(m-l)^2
    \right]    
    }
    c_{m}^*c_{l}c_{n+m-l} .
\end{equation}
with constraint $\sum_n |c_n|^2=1$. We find that numerical solutions are more stable when we scale the above equation with $g_*N$:
\begin{equation}
    \frac{\partial c_{n}}{\partial \tau} = 
    \frac{1}{g_*N}n^2 c_n+
    \sum_{m,l}
    e^{
    -2\pi^2\gamma^2
    \left[
        (n-l)^2+(m-l)^2
    \right]    
    }
    c_{m}^*c_{l}c_{n+m-l} .
    \label{eq:time-dep_GPE_rescaled}
\end{equation}
where we also redefined $\tau\rightarrow g_*N\tau$. 

We solve Eq.~\eqref{eq:time-dep_GPE_rescaled} by a standard Euler step iteration. We start with a random complex $c_n$ which is normalized and symmetrized with respect to the center of the ``lattice''. We calculate the right hand side of Eq.~\eqref{eq:time-dep_GPE_rescaled} for each $n$ as $\mathcal{H}[c_n]$. We discretize the imaginary time derivative as $\partial c_n/\partial\tau\rightarrow \left[c_n(\tau)-c_n(\tau-\delta\tau)\right]/\delta\tau$ and update the initial random coefficients by $c_n(\tau-\delta\tau)=c_n(\tau)-\delta\tau \mathcal{H}[c_n]$. After each update, we renormalize the new coefficients $c_n\rightarrow c_n/\sum_m |c_m|^2$ and calculate the chemical potential by $\mu=\sum_n c_n^*\mathcal{H}[c_n]$. Typically, $\mu$ converges after a few thousand iterations up to precision $10^{-5}-10^{-6}$. After the last iteration, we calculate all observables from $c_n$ and deduce resulting lattice structures from the analysis of them as presented below. Random initial state sometimes can get trapped in a local minimum and one needs to repeat the calculation a few times to make sure the correct ground state is found. 
We found that starting from small $g_*N$ and using the final state as an initial state for larger $g_*N$ gives better and reproducible results. Imaginary time evolution and a solution strategy related to our work is recently used to study vortex lattice structure in the symmetric gauge \cite{moroz2022}.

\begin{figure}[t]
\centering
\includegraphics[width=0.9\textwidth]{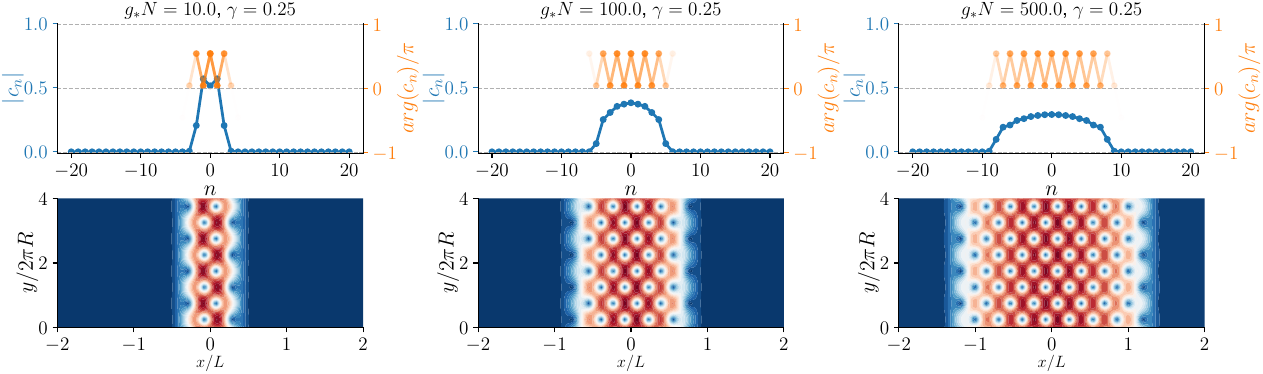}
\caption{Solution of Eq.~\eqref{eq:time-dep_GPE_rescaled} by using imaginary time evolution algorithm for different anisotropy parameters $\gamma$ and interaction constants $g_*N$. To rows show the resulting coefficients $a_n$, the absolute value in blue with scale shown in left axis, and the phase shown in orange with scale shown in right axis. Lower panels show the corresponding vortex lattice structure from $|\psi(x,y)|^2$ small values are blue and large values are red. The scale of $x$-axis and the repetition along $y$-axis are chosen manually to better display the bulk lattice structure.}
\label{figure2}
\end{figure}

In Fig.~\ref{figure2}, we show the results of imaginary time evolution for $\gamma=0.25$ for three different interactions strengths $g_*N=10$, $100$ and $500$. Note that $y$-axis is repeated about several times to better reveal the lattice structure and vortex geometry. For small $g_*N$, which may either correspond to small interaction $a_s$ or small rotation manifest in small $M_*$, we only see a few vortex rows in the system. This is deduced from a few non-zero $c_n$'s in the upper panel after the convergence albeit many more coefficients are taken in the calculation. This point is just above the well-known roton instability where the system transitions from a uniform BEC to vortex state due rotation \cite{Shlyapnikov2005}. As the rotation rate increase to $g_*N=100$, as shown in the middle panels, several vortex rows emerge with many more nonzero expansion coefficients $c_n$. At this point, cloud profile along the ``Landau level lattice'' shows an inverted parabolic shape with smooth crossover to zero values. This is reminiscent of the Thomas-Fermi profile of high density condensates but in the lowest Landau level manifold. For even larger $g_*N$, shown in the last column, more nonzero coefficients are obtained and the profile along the lattice flattens to a saturated value. In all of these points, vortex lattice shows an oblique rectangular structure broadly in agreement with experimental shots seen in \cite{ShafferMoag2021} but somewhat deformed due to boundaries along $x$-direction. Phase structure of the Landau level coefficients $c_n$ is also shown in the upper row with the orange lines and right axis labels. The phases of the coefficients seem to alternate by a phase $\pi/2$ between the even and odd sites, and interestingly there seems to be no dependence on the $n^2$ onsite potential for this parameter regime. This phase structure is broadly in agreement with the ansatz given in Ref.~\cite{Shlyapnikov2005}, but different parameters $\gamma$ and $g_*N$ yield to phases with more structure, as shown below.

\begin{figure}[t]
\centering
\includegraphics[width=8cm]{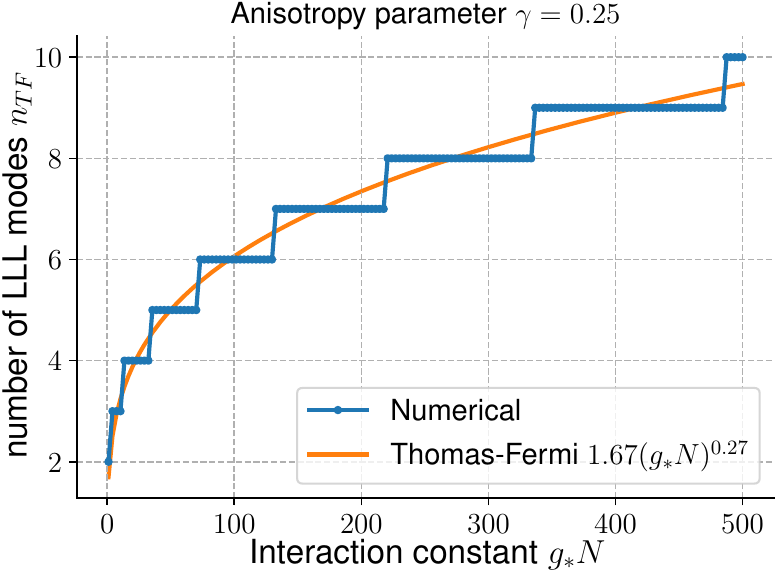}
\caption{Number of non-zero Landau level coefficients $n_{TF}$ obtained from the minimization via imaginary time evolution. More states in the Landau level activate with increasing $g_*N$ and the trend follows the Thomas-Fermi scaling, which is shown with thin (orange) line to a fit $n_{TF}=a(g_*N)^{b}$ as explained in the main text.}
\label{figure3}
\end{figure}

To pursue the anology with the Thomas-Fermi radius in the lowest Landau level, we take a dense cut along $g_*N$ axis for fixed $\gamma=0.25$ and count the number of Landau level modes $c_n$ bigger than a small threshold, say $0.01$, on the right hand side of the system, which is denoted by $n_{TF}$. As shown in Fig.~\ref{figure3}, $n_{TF}$ increases in integer steps with increasing rotation/interaction $g_*N$. A fit to the data using the generalized expression for the Thomas-Fermi radius $n_{TF}=a(g_*N)^b$ gives $a=1.67$ and $b=0.27$, which is very close to the numerical values $a=1.7$ and $b=1/5$ given in Eq.(6.56) of Ref.~\cite{pethick_smith}.

\begin{figure}[t]
\centering
\includegraphics[width=0.9\textwidth]{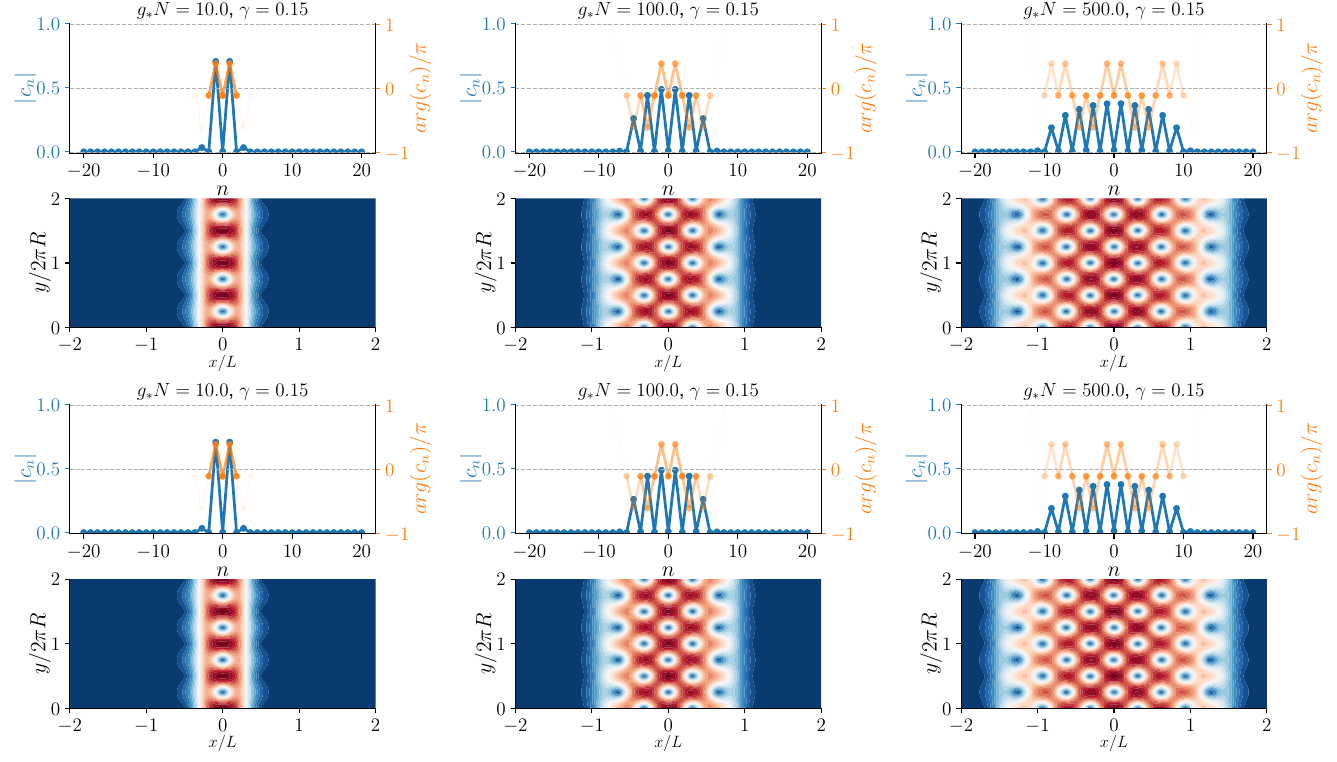}
\caption{Same as Fig.~\ref{figure2} but with different anisotropy parameters. Upper two rows are for smaller $\gamma=0.15$ and lower two rows are for bigger $\gamma=0.35$, as indicate in the titles.}
\label{figure4}
\end{figure}
Lastly, we also show the results of the minimization for different anisotropy parameters with similar cuts along $g_*N$ in Fig.~\ref{figure4}. 
For smaller anisotropy parameter, $\gamma=0.15$ shown in upper two rows, $g_*N=10$ yields only one vortex row rather than two. Different from Fig.~\ref{figure2}, the ``zeroth Landau level'' coefficient is zero $c_0=0$ here. For larger $g_*N=100$ and $g_*N=500$, we find that this continues to hold for all even coefficients, with the odd coefficient still following the Thomas-Fermi profile revealed in previous cut. The lattice structure is clearly triangular, as expected for isotropic condensates but the edges of the system deforms this slightly. Phase structure shown in thin (orange) lines are also markedly different from previous cut; they have a period of four rather than two alternating between phases approximately $\pi/2$, 0 and $-\pi/2$. 
For larger anisotropy parameter, $\gamma=0.35$ shown in lower two rows, $g_*N=10$ yields two vortex rows as in Fig.~\ref{figure2} but the vortex cores are somewhat elongated. Larger values $g_*N=100$ and $g_*N=500$, still show oblique rectangular structure but the tendency for a vortex stripe also emerge here. Landau level coefficients do not have a bi-partition and follows earlier Thomas-Fermi profile but the phase structure shows a parabolic alternation with even/odd modulation, quite different from other anisotropy parameters presented in upper rows here and Fig.~\ref{figure2}.

\section{Conclusion}
\label{section5}
In this paper we developed a theory of rapidly rotating anisotropic Bose gases. We formulated a model for rotating system in the Landau gauge, which is more suitable for anisotropic system, and considered additional perturbations breaking the translational invariance  such as the trap potential. Considering short-range interactions, we derived the many-body Hamiltonian projected to lowest Landau level manifold and derived a second quantized model with two parameters; anisotropy $\gamma$ and dimensionless interaction $g_*$. In the full Bose-Einstein condensate limit, we derived a Gross-Pitaevskii equation projected to lowest Landau level, and developed a robust solution based on imaginary time evolution algorithm. Our method was shown to be capable of finding different lattice structures in an anisotropic system. Interestingly, we revealed 
that the number of states with a support in the LLL denoted by $n_{TF}$, which determines the
number of vortices, follows a Thomas-Fermi type scaling albeit with slightly different coefficients from the usual
condensates.

There are several future directions that can be pursued from our work. Undoubtedly, the most interesting investigation would be study of full many-body system given in Eq.~\ref{eq:two_parameter_Hamiltonian}, which has the potential to expose new Fractional Quantum Hall states in the wide parameter regime of the model with interesting edge effects. Such a work requires powerful unbiased numerical schemes such as exact diagonalization of large systems or, alternatively, Density Matrix Renormalization Group method. The Gross-Pitaevskii formalism presented here can also be generalized to systems with long-range dipolar interactions and two-component gases, which has direct relevance for recent experimental systems. Last but not least, the quantum fluctuation in the form of Lee-Huang-Yang correction can be added to the lowest Landau level Gross-Pitaevskii model which can drastically modify vortex lattice structures \cite{oktel2023}.

\section*{Acknowledgment}
I thank U. Tanyeri, A. Kallushi, R. O. Umucalilar, D. Kantar, U. C. Turhan and M. O. Oktel for helpful discussions.
\\
This work is supported by 
the Scientific and Technological
Research Council of Türkiye (TUBITAK) 1001 program Project No. 124F122, and TUBITAK 2236 cofunded Brain
Circulation Scheme 2 (CoCirculation2) Project No. 120C066.

\end{document}